\documentclass{ws-ijmpb}

\begin{document}

%
%

\title{The Memory Function of the Generalized Diffusion Equation of Active Motion}
\author{FRANCISCO J. SEVILLA}

\address{Instituto de F\'{\i}sica, Universidad Nacional Aut\'{o}noma de M\'{e}xico,\\ 
Apartado Postal 20-364, 01000 Ciudad de M\'{e}xico, Mexico.\\
fjsevilla@fisica.unam.mx}

\maketitle


\begin{abstract}
An exact description of the statistical motion of active particles in three dimension is presented in the framework of a \emph{generalized diffusion equation}. Such a generalization contemplates a non-local, in time and space, \emph{connecting} (memory) function. This couples the rate of change of the probability density of finding the particle at position $\boldsymbol{x}$ at time $t$, with  the Laplacian of the probability density at all previous times and to all points in space. Starting from the standard Fokker-Planck-like equation for the probability density of finding an active particle at position $\boldsymbol{x}$ swimming along the direction $\hat{\boldsymbol{v}}$ at time $t$, we derive in this paper, in an exact manner, the connecting function that allows a description of active motion in terms of this generalized diffusion equation.          
\end{abstract}


\section{Introduction: A generalization of the diffusion equation}
Different mathematical frameworks have been ingeniously devised to describe transport phenomena, particularly, to describe the random motion of a particle suspended in an fluid in thermal equilibrium, known as Brownian motion. The connection between these apparently distinct frameworks has been fully recognized, and the transition from one to the other has been presented elsewhere. Two of these well-known frameworks refer to the \emph{diffusion equation} and the \emph{random walk}. 

Extensions of these two frameworks have been considered in order to consider non-Markovian effects, naturally associated with nonequilibrium phenomena. Regarding a generalization of the random walk framework, the \emph{continuous-time random walk} of Montroll, Weiss and West has been used in order to describe anomalously-slow diffusion, called subdiffusion, for which the variance of the displacement grows more slowly than linear in time, being the last one, the characteristic mode of diffusion of a Brownian particle. In connection with the continuous-time random walk, Professor Kenkre, with Montroll and Shlesinger, proof the equivalence with the generalized master equation\cite{KenkreJStatPhys1973,KenkrePRB1974}, which replaces the local-in-time transition rates between the system states in the standard master equation, by non-local-in-time {\it memory functions} that emerges in the passage from a microscopic description to a macroscopic one. \cite{Kenkre1977,Kenkre2003,Kenkre2021}

The contributions of V.M. (Nitant) Kenkre to non-equilibrium statistical mechanics are diverse and vast, and I picture him as an architect of overpasses, building beautiful bridges between apparently distant fields. I want to thank Nitant for the much he taught me, and with pleasure I dedicate the present article to celebrate his contributions.

In the pursuit of connecting two distinct generalizations of the diffusion equation that describe anomalous diffusion, one of such bridges was presented in Ref. \refcite{Kenkre2007} for one-dimensional systems. In there, one  generalization corresponds to the \emph{time-dependent diffusion coefficient} equation,
\begin{equation}\label{chi}
\frac{\partial }{\partial t}P_{\chi}(x,t)=D\chi(t)\frac{\partial^{2}}{\partial x^{2}}P_{\chi}(x,t),
\end{equation} 
while the other generalization is based on the the memory formalism given by
\begin{equation}\label{phi}
\frac{\partial }{\partial t}P_{\phi}(x,t)=D\int_{0}^{t}ds\,\phi(t-s)\frac{\partial^{2}}{\partial x^{2}}P_{\phi}(x,s).
\end{equation}
In general, it is not possible to find a $\chi(t)$ given $\phi(t)$ or to find $\phi(t)$ given $\chi(t)$, that makes the solutions of \eqref{chi} and \eqref{phi} equivalent, however a connection between both formalisms is possible.\cite{Kenkre2007}  

The kind of bridge needed is specific to each of the two starting points. The construction of a bridge that starts at Eq. \eqref{chi}, requires the time rate of change of $P(x,t)$ to be connected to its Laplacian not only at the spatial point $x$ at all times $0\le s\le t$, but at the all locations $x^{\prime}$ through the ``bridge roadway'' $\mathfrak{D}(x,t)$, defined by the \emph{generalized diffusion equation}
\begin{equation}
\frac{\partial}{\partial t}{P(x,t)}=\int \text{d}x^{\prime}\int_{0}^{t}\text{d}s\, \mathfrak{D}(x-x^{\prime},t-s) \frac{\partial^{2}}{\partial {x^{\prime}}^{2}}P(x^{\prime},s).
\label{genphi}
\end{equation}
The other bridge, the one that starts at \eqref{phi}, is explicitly calculated in Ref. \refcite{Kenkre2007} and it is not necessary to revisit it here, since 
Eq. \eqref{genphi} forms the basis of the present analysis.   





\section{The connecting function $\mathfrak{D}(\boldsymbol{x},t)$ of active motion}

Simple examples that make explicit the appearance of the \emph{space-time nonlocal diffusion equation} \eqref{genphi} have been given in Refs. \refcite{Kenkre2007} and \refcite{GiuggioliJPA2009}. More recently, the corresponding two-dimensional version of Eq. \eqref{genphi} has been unveiled in the context of \emph{persistent motion} of \emph{active} or \emph{self-propelled} particles\cite{SevillaPRE2020}, which maintain on the average its direction of motion by consuming energy from the environment.\cite{RomanczukEPJST2012} The diversity of patterns of active motion observed in biological organisms or in artificially designed active particles is indeed large, enriching the variety of effects exhibited by these particles.

In this paper I present an analysis, in the context of the space-time nonlocal generalized diffusion equation \eqref{genphi}, of three-dimensional active motion that follows an arbitrary, however Markovian and spatially local, orientational dynamics of the particle direction of motion or swimming direction. This dynamics is fully characterized by the probability rate for the particle swimming direction to be scattered. As pointed out by many authors, the role of spatial dimensionality is relevant, particularly in the context of rotations that underlie the scattering processes of the swimming direction considered in here. For the case considered here, we explicitly derive the connecting function $\mathfrak{D}(\boldsymbol{x},t)$, in terms of the parameters that characterizes the specific pattern of active motion.

The starting point of our analysis is the transport equation for the probability density, $P({\boldsymbol{x}},\hat{\boldsymbol{v}} ,t)$, of a single particle being at position $\boldsymbol{x}$, moving at constant speed $v_{0}$ along the swimming direction $\hat{\boldsymbol{v}}$ at time $t$, that is,
\begin{equation}\label{TransportEquation}
\frac{\partial }{\partial t}P({\boldsymbol{x}},\hat{\boldsymbol{v}} ,t)+v_{0}\hat{\boldsymbol{v}}\cdot \nabla P({\boldsymbol{x}},\hat{\boldsymbol{v}}
,t)= \int \text{d}\hat{\boldsymbol{v}}^{\prime}\,  K_A\left(\hat{\boldsymbol{v}}\vert\hat{\boldsymbol{v}}^{\prime}\right)P(\boldsymbol{x},\hat{\boldsymbol{v}}^{\prime},t),
\end{equation}
where $K_A\left(\hat{\boldsymbol{v}}\vert\hat{\boldsymbol{v}}^{\prime}\right)$ gives the probability rate of the transition from the swimming direction $\hat{\boldsymbol{v}}^{\prime}$ to $\hat{\boldsymbol{v}}$, and encompasses the detailed information of a specific pattern of active motion considered. Last equation 
is some times called the \emph{active-transport equation} 
which can be rewritten in the gain-loss form, 
\begin{multline}\label{TransportEquation2}
\frac{\partial }{\partial t}P({\boldsymbol{x}},\hat{\boldsymbol{v}} ,t)+v_{0}\hat{\boldsymbol{v}}\cdot \nabla P({\boldsymbol{x}},\hat{\boldsymbol{v}} ,t)=\int\text{d}\hat{\boldsymbol{v}}^{\prime}\, Q\left(\hat{\boldsymbol{v}},\hat{\boldsymbol{v}}^{\prime}\right)P(\boldsymbol{x},\hat{\boldsymbol{v}}^{\prime},t)\\
-\left[\int\text{d}\hat{\boldsymbol{v}}^{\prime} Q\left(\hat{\boldsymbol{v}}^{\prime},\hat{\boldsymbol{v}}\right)\right]P(\boldsymbol{x},\hat{\boldsymbol{v}},t),
\end{multline}
when $K_A\left(\hat{\boldsymbol{v}}\vert\hat{\boldsymbol{v}}^{\prime}\right)$ is written in terms of the distribution of scattering $Q(\hat{\boldsymbol{v}},\hat{\boldsymbol{v}}^{\prime})$ between the swimming directions $\hat{\boldsymbol{v}}$ and $\hat{\boldsymbol{v}}^{\prime}$, which inherits from $K_A\left(\hat{\boldsymbol{v}}\vert\hat{\boldsymbol{v}}^{\prime}\right)$ the main characteristics of active motion, namely
\begin{equation}\label{GainLossTransitions}
 K_{A}(\hat{\boldsymbol{v}}\vert\hat{\boldsymbol{v}}^{\prime})=Q(\hat{\boldsymbol{v}},\hat{\boldsymbol{v}}^{\prime})-\delta(\hat{\boldsymbol{v}}-\hat{\boldsymbol{v}}^{\prime})\int \text{d}\hat{\boldsymbol{v}}^{\prime\prime}Q(\hat{\boldsymbol{v}}^{\prime\prime},\hat{\boldsymbol{v}}).
\end{equation}
It is intuitively clear that the sum over all directions $\hat{\boldsymbol{v}}^{\prime}$ of $Q(\hat{\boldsymbol{v}},\hat{\boldsymbol{v}}^{\prime})$ while keeping $\hat{\boldsymbol{v}}$ fixed, is rotationally invariant, i.e., it is a constant, lets say $\Lambda$, that does not depend on $\hat{\boldsymbol{v}}$.\footnote{A suitable rotation of the a Euclidean coordinates system can make to coincide $\hat{\boldsymbol{v}}$  with the $z$-axis, thus the sum over all $\hat{\boldsymbol{v}}^{\prime}$ has the same value for any direction of $\hat{\boldsymbol{v}}$.}
$\Lambda$ gives the inverse of the timescale that measures the average time between transitions, therefore
\begin{multline}\label{TransportEquation3}
\frac{\partial }{\partial t}P({\boldsymbol{x}},\hat{\boldsymbol{v}} ,t)+v_{0}\hat{\boldsymbol{v}}\cdot \nabla P({\boldsymbol{x}},\hat{\boldsymbol{v}},t)=\Lambda\int\text{d}\hat{\boldsymbol{v}}^{\prime} \widetilde{Q}\left(\hat{\boldsymbol{v}},\hat{\boldsymbol{v}}^{\prime}\right)P(\boldsymbol{x},\hat{\boldsymbol{v}}^{\prime},t)
-\Lambda P(\boldsymbol{x},\hat{\boldsymbol{v}},t),
\end{multline}
where the normalized scattering function $\widetilde{Q}(\hat{\boldsymbol{v}},\hat{\boldsymbol{v}}^{\prime})=Q(\hat{\boldsymbol{v}},\hat{\boldsymbol{v}}^{\prime})/\Lambda$ has been introduced.

Given the linear nature of Eq. \eqref{TransportEquation3} and of the spatial homogeneity of the process, the search for a solution simplifies by transforming the position variables to Fourier ones, i.e., 
\begin{multline}\label{TransportEquationFourier}
\frac{\partial }{\partial t}\widetilde{P}({\boldsymbol{k}},\hat{\boldsymbol{v}} ,t)+\text{i}\, v_{0}\hat{\boldsymbol{v}}\cdot \boldsymbol{k}\,\, \widetilde{P}({\boldsymbol{k}},\hat{\boldsymbol{v}},t)=\Lambda\int\text{d}\hat{\boldsymbol{v}}^{\prime} \widetilde{Q}\left(\hat{\boldsymbol{v}},\hat{\boldsymbol{v}}^{\prime}\right)\widetilde{P}(\boldsymbol{k},\hat{\boldsymbol{v}}^{\prime},t)-\Lambda \widetilde{P}(\boldsymbol{k},\hat{\boldsymbol{v}},t),
\end{multline}
where $\widetilde{P}(\boldsymbol{k},\hat{\boldsymbol{v}},t)=\int d\boldsymbol{x}\, e^{-\text{i}\boldsymbol{k}\cdot\boldsymbol{x}}P(\boldsymbol{x},\hat{\boldsymbol{v}},t)$. In addition, given that the unit vector $\hat{\boldsymbol{v}}$ marks a point on the unit sphere $\mathbb{S}^{2}$ embedded in the three-dimensional space, this can be parametrized by the standard spherical coordinates $\theta$ and $\varphi$. Thus, we expand $\widetilde{P}({\boldsymbol{k}},\hat{\boldsymbol{v}} ,t)$ in the set of the scalar spherical harmonics on $\mathbb{S}^{2}$ given by
\begin{equation}\label{SpherHar}
Y_{l}^{m}(\hat{\boldsymbol{v}})\equiv Y_{l}^{m}(\theta,\varphi)=(-1)^{m}\sqrt{\frac{(2l+1)}{4\pi}\frac{(n-m)!}{(n+m)!}}  \, \text{P}_{l}^{m}(\cos\theta) e^{\text{i}m\varphi},
\end{equation}
as:
\begin{equation}
\widetilde{P}({\boldsymbol{k}},\hat{\boldsymbol{v}} ,t)=\sum_{l,m}\bigl[\widetilde{p}_{l}^{m}(\boldsymbol{k},t)\, e^{-\lambda_{l}^{m}t}\bigr]\, Y_{l}^{m}(\hat{\boldsymbol{v}}),
\end{equation}
where the expansion coefficients 
\begin{equation}\label{CoeffExpansion}
\widetilde{p}_{l}^{m}(\boldsymbol{k},t)=e^{\lambda_{l}^{m}t}\int\text{d}\hat{\boldsymbol{v}}\, {Y^{*}}_{l}^{m}(\hat{\boldsymbol{v}})\widetilde{P}({\boldsymbol{k}},\hat{\boldsymbol{v}} ,t)
\end{equation}
are to be determined. $\text{P}_{l}^{m}(x)$ are the associated Lengedre functions and the symbol $\sum_{lm}$ denotes $\sum_{l=0}^{\infty}\sum_{m=-l}^{l}$.
The factors $e^{-\lambda_{l}^{m}t}$ correspond to the coefficients of the expansion $f(\hat{\boldsymbol{v}},t)=\sum_{l,m} e^{-\lambda_{l}^{m}t}\, Y_{l}^{m}(\hat{\boldsymbol{v}})$ that solves the equation
\begin{equation}\label{Lambdas}
\frac{\partial }{\partial t}f(\hat{\boldsymbol{v}} ,t)=\Lambda\int\text{d}\hat{\boldsymbol{v}}^{\prime} \widetilde{Q}\left(\hat{\boldsymbol{v}},\hat{\boldsymbol{v}}^{\prime}\right)f(\hat{\boldsymbol{v}}^{\prime},t)-\Lambda f(\hat{\boldsymbol{v}},t).
\end{equation}
From Eq. \eqref{CoeffExpansion} we have that in position variables $\boldsymbol{x}$
\begin{equation}
p_{l}^{m}(\boldsymbol{x},t)=e^{\lambda_{l}^{m}t}\int\text{d}\hat{\boldsymbol{v}}\, {Y^{*}}_{l}^{m}(\hat{\boldsymbol{v}})P({\boldsymbol{x}},\hat{\boldsymbol{v}} ,t).
\end{equation}

Progress is made by considering that the scattering weight $\widetilde{Q}(\hat{\boldsymbol{v}},\hat{\boldsymbol{v}}^\prime)$ depends only on the relative angle between $\hat{\boldsymbol{v}}$ and $\hat{\boldsymbol{v}}^{\prime}$, $\vartheta$. In such a case, the scattering process consists of a rotation around any of the directions $\pm\bigl(\hat{\boldsymbol{v}}\times\hat{\boldsymbol{v}}^{\prime}\bigr)$ by a randomly chosen angle $\vartheta$ in $[0,\pi]$\footnote{The process considered here scatters the direction $\hat{\boldsymbol{v}}^{\prime}$ into $\hat{\boldsymbol{v}}^{\prime}$ through a geodesic that joins both vectors on the two-dimensional sphere $\mathbb{S}^{2}$.}, thus $\widetilde{Q}(\hat{\boldsymbol{v}},\hat{\boldsymbol{v}}^{\prime})=\widetilde{Q}(\hat{\boldsymbol{v}}\cdot\hat{\boldsymbol{v}}^{\prime})=\widetilde{Q}(\cos\vartheta)$. From these considerations we have that 
\begin{equation}
\lambda_{l}^{m}=\lambda_{l}=\Lambda\Bigl[1-\bigl\langle\text{P}_{l}(\hat{\boldsymbol{v}}\cdot\hat{\boldsymbol{v}}^{\prime})\bigr\rangle_{\widetilde{Q}}\Bigr],
\end{equation}
where 
\begin{equation}
\bigl\langle\text{P}_{l}(\hat{\boldsymbol{v}}\cdot\hat{\boldsymbol{v}}^{\prime})\bigr\rangle_{\widetilde{Q}}=2\pi\int_{0}^{\pi}d\vartheta\sin\vartheta\, \widetilde{Q}(\cos\vartheta)\text{P}_{l}(\cos\vartheta).
\end{equation}
Since the Legendre functions are bounded, it can be proved that $0\le\lambda_{l}\le2\Lambda$ and it is easy to show that $\lambda_{l=0}=0$. 
\begin{figure}[t]
\centerline{\psfig{file=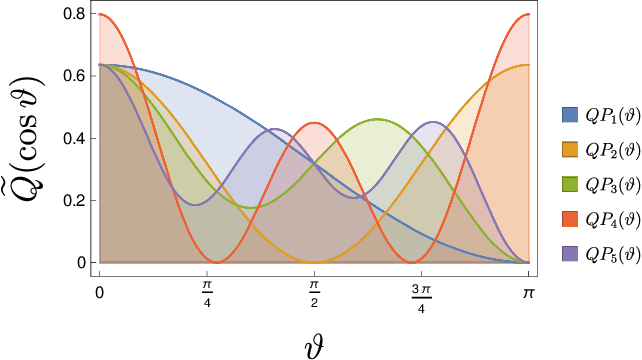,width=0.75\textwidth}}
{\vspace*{8pt}}
\centerline{\psfig{file=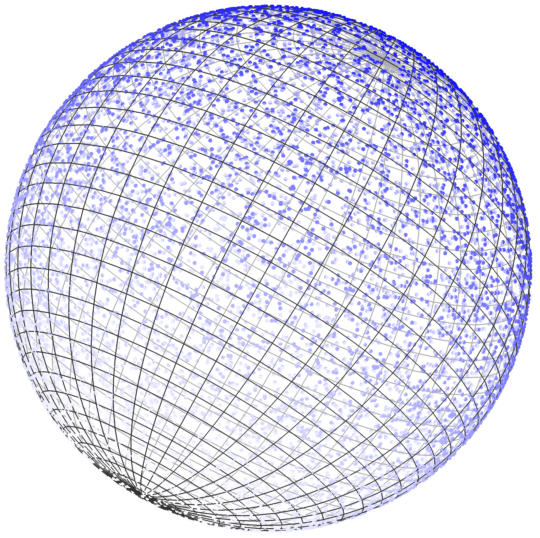,width=0.2\textwidth}\psfig{file=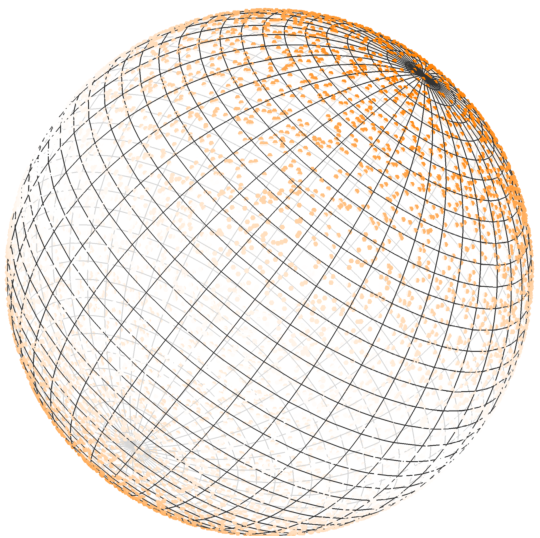,width=0.2\textwidth}\psfig{file=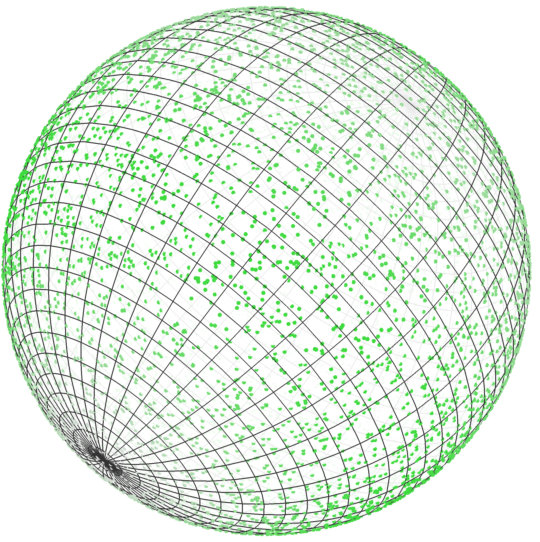,width=0.2\textwidth}\psfig{file=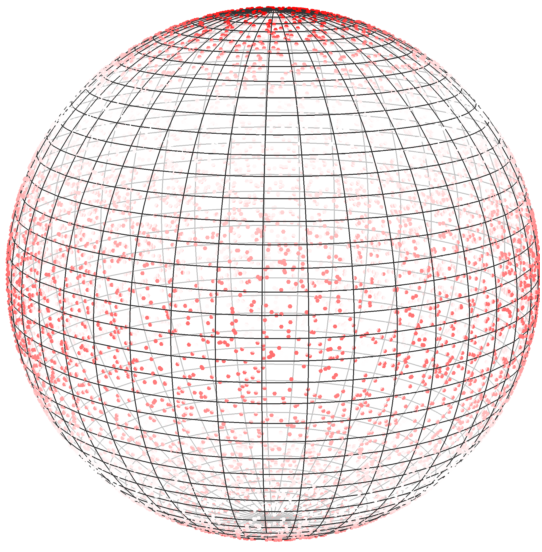,width=0.2\textwidth}\psfig{file=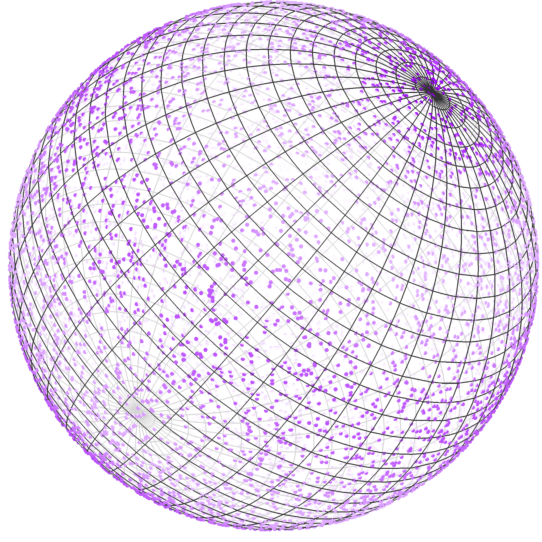,width=0.2\textwidth}}
 {\vspace*{8pt}}\caption{\label{QLfunctions} The scattering function $\widetilde{Q}(\cos\vartheta)$ is shown as function of the scattering angle $\vartheta$. Top panel.- Different scattering functions, given by $QP_{l}(\vartheta)$ $l=0,1,\ldots$, implement different swimmers navigation strategies: Active Brownian motion (blue line), run-and-reverse (orange line), run-and-flick (green line), run-flick-and-reverse (red line) and run-flick-and-flick (purple). In the bottom panel, the directions $\hat{\boldsymbol{v}}^{\prime}$ in Eq. \eqref{QPl} that scatter into $\hat{\boldsymbol{v}}$ in Eq. \eqref{QPl} are sampled according to $QP_{l}(\vartheta)$ and shown on the unitary sphere, from left to right $l=1$ (active Brownian motion), 2 (run-and-reverse), 3 (run-and-flick), 4 (run-flick-and-reverse) and 5 (run-flick-and-flick).}
\end{figure}
Different navigation strategies of the swimmer are encompassed in this last quantity. Here I consider specific cases of scattering functions that represent typical navigation strategies. These are modeled as modifications of the Legendre functions $\text{P}_{l}(\cos\vartheta)$. Namely we choose
\begin{equation}\label{QPl}
QP_{l}(\vartheta)=\frac{1}{N_{l}}[\text{P}_{l}(\cos\vartheta)+c_{l}],
\end{equation}
where $c_{l}$ is a constant that shifts the Legendre functions such that $QP_{l}(\vartheta)\ge0$ and $N_{l}$ is a normalization constant. The first six $QP_{l}(\vartheta)$ functions are given in the Table \ref{tab1}, while the corresponding ones $l=0$, 1, 2, 3, 4 and 5  are shown as function of $\vartheta$ in the top panel of Figure \ref{QLfunctions}.

The simplest navigation strategy, the so-called \emph{run-and-tumble} motion, corresponds to the case for which scattering to any direction is equally probable, i.e. $\widetilde{Q}(\hat{\boldsymbol{v}}\cdot\hat{\boldsymbol{v}}^{\prime})=1/4\pi$, for which we have $\lambda_{l}=\Lambda[1-\delta_{l,0}]$. On the other hand, the class of functions $\widetilde{Q}(\cos\vartheta)$ that promotes scattering in the forward direction only, are qualitatively close to the so called active Brownian motion, and are represented by $QP_{l}(\vartheta)$ with $l=1$ (see Fig. \ref{QLfunctions}, blue line in the top panel and first sphere from the left in the bottom panel). The case $l=2$ (orange in top panel second sphere from the left in the bottom panel) describes scattering processes that occur predominantly around the forward and backward directions, which model in a simplistic way the \emph{run modes} of microorganisms that exhibit run-and-reverse motion.\cite{ThevesBioPhysJ2013} Other simple scattering functions can describe different swimming strategies, some of them are given in Table~\ref{tab1}.  
\begin{table}[h]   
\tbl{The first six scattering functions, $QP_{l}(\vartheta)$, $l=0,$ 1, 2, 3, 4 and 5 used in this paper and their corresponding pattern of active motion they describe.}
{\begin{tabular}{@{}ccl@{}} \Hline 
\\[-1.8ex] 
$l$ & $QP_{l}(\vartheta)$ & Pattern of motion \\[0.8ex] 
\hline \\[-1.8ex] 
$0$ & $\frac{1}{2}$ & Run-and-tumble\\
$1$ & $\frac{1}{\pi}(\cos\vartheta+1)$ & Active Brownian\\
$2$ & $\frac{1}{\pi}\bigl(\cos(2\vartheta)+1\bigr)$  & Run-and-reverse\\ 
$3$ & $\frac{1}{8\pi}\bigl(3\cos\vartheta+5\cos(3\vartheta)\bigr)+\frac{1}{\pi}$ & Run-and-flick \\ 
$4$ & $\frac{2}{51\pi}\bigl(1+7\cos(2\vartheta)\bigr)^{2}$ & Run-flick-and-reverse \\ 
$5$ & $\frac{1}{8\pi}\bigl(63\cos^{5}\vartheta-70\cos^{3}\vartheta+15\cos\vartheta+8\bigr)$ &  Run-flick-flcik
\\[0.8ex] 
\Hline \\[-1.8ex] 
\multicolumn{3}{@{}l}{
}\\
\end{tabular}}
\label{tab1}
\end{table}

With these considerations we have that the rate of change of the expansion coefficients $\widetilde{p}_{l}^{m}(\boldsymbol{k},t)$ are given by 
\begin{equation}\label{SphHarmCoefficients}
 \frac{d}{dt}\widetilde{p}_{l}^{m}(\boldsymbol{k},t)=-\text{i}v_{0}\sum_{l^{\prime},m^{\prime}}\widetilde{p}_{l^{\prime}}^{m^{\prime}}(\boldsymbol{k},t)\, e^{-(\lambda_{l^{\prime}}-\lambda_{l})t}\int d\hat{\boldsymbol{v}}\, {Y}_{l}^{m*}(\hat{\boldsymbol{v}})\, \left[\hat{\boldsymbol{v}}\cdot\boldsymbol{k}\right]\, Y_{l^{\prime}}^{m^{\prime}}(\hat{\boldsymbol{v}}),
\end{equation}
where the orthogonality property of the spherical harmonics has been used. The integral in \eqref{SphHarmCoefficients} gives the explicit coupling factors among the coefficients $\widetilde{p}_{l}^{m}(\boldsymbol{k},t)$ owing to the advection term related to self-propulsion, $\text{i}v_{0}\hat{\boldsymbol{v}}\cdot\boldsymbol{k}$, in Eq.~\eqref{TransportEquationFourier}. Due to the rotational invariance of the integral in \eqref{SphHarmCoefficients} we can choose coordinates, $\hat{\boldsymbol{v}}^{\prime},$ such the polar axis points along the direction of $\boldsymbol{k}$. In such a case 
\begin{align}
\int d\hat{\boldsymbol{v}}\, {Y}_{l}^{m*}(\hat{\boldsymbol{v}})\, \left[\hat{\boldsymbol{v}}\cdot\boldsymbol{k}\right]\, Y_{l^{\prime}}^{m^{\prime}}(\hat{\boldsymbol{v}})&=k\int d\hat{\boldsymbol{v}}^{\prime}\, {Y}_{l}^{m*}(\hat{\boldsymbol{v}}^{\prime})\, \cos\vartheta^{\prime}\, Y_{l^{\prime}}^{m^{\prime}}(\hat{\boldsymbol{v}}^{\prime}),\\
&=k\, I_{l,l^{\prime}}^{m,m^{\prime}},
\end{align}
where $\vartheta^{\prime}$ is the angle between $\hat{\boldsymbol{v}}^{\prime}$ and $\boldsymbol{k}$ and
\begin{multline}
  I^{m,m^{\prime}}_{l,l^{\prime}}=\delta_{l,l^{\prime}+1}\delta_{m,m^{\prime}}\left[\frac{(l^{\prime}-m^{\prime}+1)(l^{\prime}+m^{\prime}+1)}{(2l^{\prime}+1)(2l^{\prime}+3)}\right]^{1/2}+\\\delta_{l,l^{\prime}-1}\delta_{m,m^{\prime}}\left[\frac{(l^{\prime}-m^{\prime})(l^{\prime}+m^{\prime})}{(2l^{\prime}-1)(2l^{\prime}+1)}\right]^{1/2}.
\end{multline}

Equation \eqref{SphHarmCoefficients} can be solved by use of the Laplace transform, for the initial condition $\widetilde{p}_{l}^{m}(\boldsymbol{k})=\delta_{l,0}\delta_{m,0}/4\pi$. This corresponds to the case for which an ensemble of particles start at the origin of coordinates, with orientations equally distributed on the whole solid angle $4\pi$. The notation here must be understood as follows: we write simply $\widetilde{p}_{l}^{m}(\boldsymbol{k})$ to denote the Fourier transform of the distribution $p_{l}^{m}(\boldsymbol{x},t)$ at time $t=0$, and we write $\widetilde{p}_{l}^{m}(\boldsymbol{k},\epsilon)$ to denote the Laplace transform of $\widetilde{p}_{l}^{m}(\boldsymbol{k},t)$, i.e., the second argument, $t$ or $\epsilon$, distinguishes the description in time or Laplace domain, respectively. Thus, after Laplace transforming \eqref{SphHarmCoefficients} and after some straightforwardly algebraic steps we get the following hierarchy of equations
\begin{multline}\label{LapSphHarmCoefficients}
 \widetilde{p}_{l}^{m}(\boldsymbol{k},\epsilon)=\frac{1}{\epsilon}\widetilde{p}_{l}^{m}(\boldsymbol{k})-\text{i}\frac{v_{0}k}{\epsilon}\Biggl[\sqrt{\frac{(l-m)(l+m)}{(2l-1)(2l+1)}}\;\; \widetilde{p}_{l-1}^{m}(\boldsymbol{k},\epsilon+\lambda_{l-1}-\lambda_{l})\\
+\sqrt{\frac{(l-m+1)(l+m+1)}{(2l+1)(2l+3)}}\;\; \widetilde{p}_{l+1}^{m}(\boldsymbol{k},\epsilon+\lambda_{l+1}-\lambda_{l})\Biggr].
\end{multline}
From which we notice that the coefficient $\widetilde{p}_{l}^{m}(\boldsymbol{k},\epsilon)$ is coupled to the corresponding ones with the same $m$ but with coefficients differing from $l$ by $\pm1$ and the Laplace variable shifted by $\lambda_{l\pm1}-\lambda_l$, i.e., $\epsilon\rightarrow\epsilon+\lambda_{l\pm1}-\lambda_l$. 

\subsection{The generalized diffusion equation of active motion}
If the information contained in the probability density $P(\boldsymbol{x},\hat{\boldsymbol{v}},t)$ is reduced in such a way that we focus our analysis on $P(\boldsymbol{x},t)=\int d\hat{\boldsymbol{v}}\, P(\boldsymbol{x},\hat{\boldsymbol{v}},t)$, we ask if the generalized diffusion equation \eqref{genphi} properly describes the motion of an active particle whose navigation strategy is defined by the scattering function $\widetilde{Q}$, and if possible, what is the relation between the memory function $\mathfrak{D}(\boldsymbol{x},t)$ and $\widetilde{Q}(\hat{\boldsymbol{v}},\hat{\boldsymbol{v}}^{\prime})$? First note that $P(\boldsymbol{x},t)=\sqrt{4\pi}\, p_{0}^{0}(\boldsymbol{x},t)$, and that $p_{0}^{0}(\boldsymbol{x},t)$ can be obtained after Laplace and Fourier inversion of $\widetilde{p}_{0}^{0}(\boldsymbol{k},\epsilon)$, which is obtained from the coupled equations \eqref{LapSphHarmCoefficients}. The result for this last quantity is
\begin{equation}\label{ContFracSolution}
\widetilde{p}_{0}^{0}(\boldsymbol{k},\epsilon)=\cfrac{\widetilde{p}_{0}^{0}(\boldsymbol{k})}{\epsilon
	  + \frac{1}{3}\cfrac{v_{0}^{2}k^{2}}{\epsilon+\lambda_{1}
          + \frac{2}{15}\cfrac{v_{0}^{2}k^{2}}{\epsilon+\lambda_{2}
          + \frac{9}{35}\cfrac{v_{0}^{2}k^{2}}{\epsilon+\lambda_{3}
	  + \frac{16}{63}\cfrac{v_{0}^{2}k^{2}}{\epsilon+\lambda_{4}
	  + \frac{25}{99}\cfrac{v_{0}^{2}k^{2}}{\epsilon+\lambda_{5}+\ddots
           } } }}}
           },
\end{equation}
which can be rewritten as
\begin{equation}\label{GenDiffSolution}
\widetilde{p}_{0}^{0}(\boldsymbol{k},\epsilon)=\frac{\widetilde{p}_{0}^{0}(\boldsymbol{k})}{\epsilon+k^{2}\widetilde{\mathfrak{D}}(\boldsymbol{k},\epsilon)}.
\end{equation}
This corresponds exactly to the Fourier-Laplace transform of the three-dimensional version of the diffusion equation \eqref{genphi}, where the connecting function $\widetilde{\mathfrak{D}}(\boldsymbol{k},\epsilon)$ is explicitly given by
\begin{equation}\label{ConectingFunction}
\widetilde{\mathfrak{D}}(\boldsymbol{k},\epsilon)=\frac{v_{0}^{2}}{3}\cfrac{1}{
	   \epsilon+\lambda_{1}
          + \frac{2}{15}\cfrac{v_{0}^{2}k^{2}}{\epsilon+\lambda_{2}
          + \frac{9}{35}\cfrac{v_{0}^{2}k^{2}}{\epsilon+\lambda_{3}
	  + \frac{16}{63}\cfrac{v_{0}^{2}k^{2}}{\epsilon+\lambda_{4}
	  + \frac{25}{99}\cfrac{v_{0}^{2}k^{2}}{\epsilon+\lambda_{5}+\ddots
           } } } } 
           },
\end{equation}
where the characteristic navigation strategy of active motion, defined by the scattering function, is inherited in the quantities $\lambda_{l}$.   

General properties can be deduced from the present analysis. In the following we present some aspects in the long and short-time regimes.

\paragraph*{The diffusive limit: Long-time and large-length regime}
The zero-th approximant, $\widetilde{\mathfrak{D}}^{(0)}(\boldsymbol{k},\epsilon)$, of $\widetilde{\mathfrak{D}}(\boldsymbol{k},\epsilon)$, valid in the long-time and large-length regime, is $\boldsymbol{k}$ and $\epsilon$ independent, which gives rise to the local, in time and position, memory function $\mathfrak{D}(\boldsymbol{x},t)=\mathfrak{D}_\text{eff}\, \delta(\boldsymbol{x})\delta(t)$, where the effective diffusion constant $\mathfrak{D}_\text{eff}$ is given by 
\begin{equation}\label{ConectingFunction0}
\widetilde{\mathfrak{D}}^{(0)}(\boldsymbol{k},\epsilon)=\mathfrak{D}_\text{eff}=\frac{1}{\lambda_{1}}\biggl(\frac{v_{0}}{\sqrt{3}}\biggr)^{2}.
\end{equation}
In this expression $\lambda_{1}$ corresponds to the inverse of the persistence time of self-propulsion, and $v_{0}/\sqrt{3}$ is the effective speed. The persistence length $\ell_\text{p}$, is introduced through the relation $v_{0}\lambda_{1}$. Since $0\le \lambda_{1}\le 2\Lambda$, the effective diffusion is bounded from below, i.e. $\frac{1}{\Lambda}\frac{v_{0}^{2}}{6}\le\mathfrak{D}_\text{eff}$. In this regime, $p_{0}^{0}(\boldsymbol{x},t)$ is asymptotically given by the Gaussian $\exp\{-x^{2}/4\mathfrak{D}_\text{eff}t\}/(4\pi\mathfrak{D}_\text{eff}t)^{3/2}$ and the mean squared displacement is given by $\langle\boldsymbol{x}^{2}(t)\rangle\sim 6\mathfrak{D}_\text{eff}t$.

The immediate correction to the previous results, takes into account the dependence on $\boldsymbol{k}$ and $\epsilon$ of the first approximant,
\begin{equation}\label{ConectingFunction1}
\widetilde{\mathfrak{D}}^{(1)}(\boldsymbol{k},\epsilon)=\frac{v_{0}^{2}}{3}\frac{1}{\epsilon+\lambda_{1}+\dfrac{2}{15}\dfrac{v_{0}^{2}k^{2}}{\lambda_{2}}},
\end{equation}
which leads to the Fourier space connecting function
\begin{equation}\label{ConectingFunction1time}
\widetilde{\mathfrak{D}}^{(1)}(\boldsymbol{k},t)=\frac{v_{0}^{2}}{3}e^{-\lambda_{1}t}\exp\biggl\{-R^{2}(t)k^{2}\biggr\}
\end{equation}
with $R^{2}(t)\equiv\mathfrak{D}_\text{R}t,$ is the range of influence of the connecting function, and $\mathfrak{D}_\text{R}\equiv\frac{2}{15}\frac{v_{0}^{2}}{\lambda_{2}}$ being a diffusion-constant-like quantity, involving $\lambda_{2}$, that describes the diffusion of the influence of a point in space $\boldsymbol{x}$ to another $\boldsymbol{x}^{\prime}$ in the time interval $t-t^{\prime}$, as can appreciated after inversion of the Fourier transform, namely
\begin{equation}\label{ConectingFunction1timespace}
\mathfrak{D}^{(1)}(\boldsymbol{x}-\boldsymbol{x}^{\prime},t-t^{\prime})=\frac{v_{0}^{2}}{3}e^{-\lambda_{1}(t-t^{\prime})}\frac{\exp\Bigl\{-\frac{(\boldsymbol{x}-\boldsymbol{x}^{\prime})^{2}}{4R^{2}(t-t^{\prime})}\Bigr\}}{\bigl[4\pi R^{2}(t-t^{\prime})\bigr]^{3/2}}.
\end{equation}
 
In the limit $k\rightarrow0$ (large-length regime), the connecting \eqref{ConectingFunction1time} reduces the time memory function that leads to the well-known telegrapher's equation.

\paragraph*{The wave-like limit: Short-time and small-length regime}
Contrary to the previous case, the role of the memory function in the solution \eqref{GenDiffSolution} is definite in the short-time regime. In this regime, the time scales $\lambda_{l}^{-1}$ are negligible and we have that the Laplace inversion of Eq. \eqref{ContFracSolution} can be approximated by the series expansion
\begin{equation}\label{ShortTime}
\widetilde{p}_{0}^{0}(\boldsymbol{k},t)=\widetilde{p}_{0}^{0}(\boldsymbol{k})\biggl[1-\frac{1}{3!}(kv_{0}t)^{2}+\frac{2}{3^{2}}\frac{1}{5!}(kv_{0}t)^{4}-\frac{2}{5^{2}}\frac{1}{7!}(kv_{0}t)^{6}+\ldots\biggr].
\end{equation}
As is expected, the effects of the persistence of active motion are conspicuously revealed in the short-time regime. Particularly, the initial distribution of the particle positions considered here (a pulse localized at the origin), evolves into a spherical shell front of small, but finite, width. In the limit $kv_{0}t\ll1$, such a distribution is indistinguishable from the spherical shell $\delta\bigl(\vert\boldsymbol{x}\vert-v_{0}t\bigr)/4\pi\vert\boldsymbol{x}\vert^{2}$, which corresponds to the expression $\widetilde{p}_{0}^{0}(\boldsymbol{k},t)=j_{0}(kv_{0}t)$ in Fourier space, whose first two terms coincide with first two terms of the series expansion \eqref{ShortTime}, $j_{0}(x)=\sin x /x$ being the zeroth spherical Bessel function.

\begin{figure}[t]
\centerline{\psfig{file=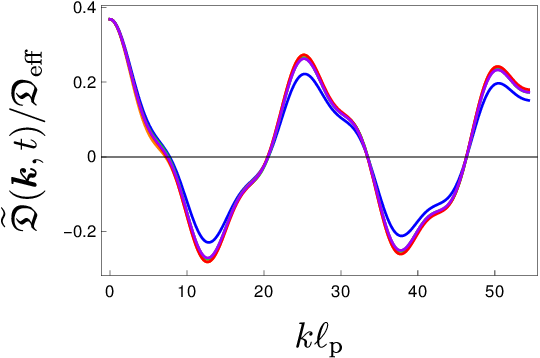,width=0.42\textwidth}\psfig{file=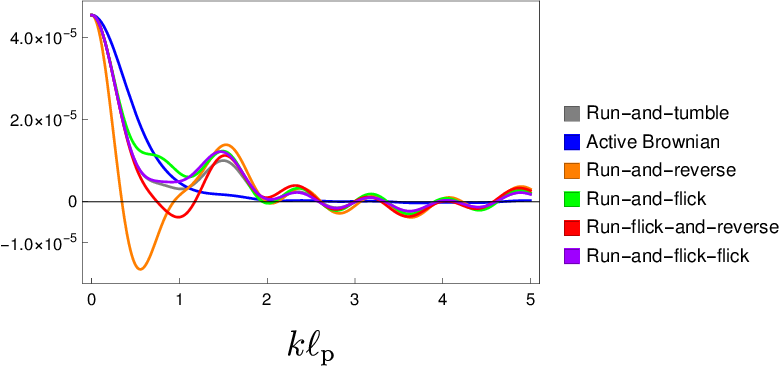,width=0.58\textwidth}}
 {\vspace*{8pt}}\caption{\label{ConnectingD} The dimensionless connecting function $\widetilde{\mathfrak{D}}(\boldsymbol{k},t)/\mathfrak{D}_\text{eff}$ is shown as function of the dimensionless wave-vector $k\ell_\text{p}$ at $\lambda t=1$ (left panel) and $\lambda t=10$ (right panel). The color indicates the corresponding pattern of active motion.}
\end{figure}

Finally, in Fig. \ref{ConnectingD}, the Fourier transform of the connecting function $\mathfrak{D}(\boldsymbol{x},t)$ is shown as function of the dimensionless wave-vector $k\ell_\text{p}$ for the distinct patterns of active motion models given in table~\ref{tab1} (see color code in Fig.~\ref{tab1}), at the dimensionless time $t\lambda_{1}=1$ (for which the persistent effects are still present, left panel) and $t\lambda_{1}=10$ (for which the connecting function is damped-out for large $k\ell_\text{p}$, right panel).

\section{Concluding remarks}
I have presented an analysis of active motion in three dimensions. Such analysis considers an arbitrary stochastic dynamics of the swimming direction that models diverse patterns of active motion, this makes our analysis to have potential applications in the analysis of the diffusion of self-propelled particles, either biological or synthetic. The analysis was carried out in the context of a \emph{generalized}, non-local in space and time, \emph{diffusion equation} formerly introduced by professor Kenkre and later, by Kenkre and the author of the present contribution, in the study of anomalous diffusion. In this paper an explicit solution of the active-transport equation is presented and the corresponding connecting function as well, this last one encompassing the relevant scales of active motion. 

Given that the time scale that characterizes the scattering process of the particle swimming direction is finite,  $\Lambda^{-1}$, the connecting function in the generalized diffusion equation evolves from a tangled space-time function, into a space-time separable one in the long-time regime. It is of interest to extend the present study to the case for which the connecting function stays tangled in time and space, particularly in the physically relevant situation for which the influence between two points in space is limited by a finite speed.

\section*{Acknowledgements}
I gladly thank Nitant Kenkre for his kind hospitality, both scientific and humanistic, while visiting the Consortium of the Americas for the Interdisciplinary Sciences from 2004 to 2008. This work was supported by UNAM-PAPIIT IN110120.


\begin{thebibliography}{0}

\bibitem{KenkreJStatPhys1973} V. M. Kenkre, E. W. Montroll, and M. F. Shlesinger, {\it J. Stat. Phys} {\bf 9}, 45 (1973).
\bibitem{KenkrePRB1974} V.M. Kenkre,  and R.S. Knox, 
Phys. Rev. {\bf B9}, 5279 (1974).

\bibitem{Kenkre1977} V.M. Kenkre, "The generalized master equation and its applications", in \emph{Statistical mechanics and statistical methods in theory and application} ed. U. Landman (New York: Plenum, 1977),  pp. 441-461.

\bibitem{Kenkre2003}V. M. Kenkre, Memory Formalism, Nonlinear Techniques, and Kinetic Equation Approaches, in \emph{AIP Conference Proceedings on Modern Challenges
in Statistical Mechanics: Patterns, Noise, and the Interplay of Nonlinearity and Complexity}, eds. V. M. Kenkre and K. Lindenberg (Melville, NY: American Institute of Physics, 2003), Vol. 658, pp. 63-103.

\bibitem{Kenkre2021} V.M. (Nitant) Kenkre: Memory Functions, Projection Operators, and the Defect Technique: Tools of the Trade for the Condensed Matter Physicist, Springer Nature Publishing.

\bibitem{Kenkre2007} V.M. Kenkre and F. J. Sevilla, in Contributions to Mathematical Physics: a Tribute to Gerard G. Emch TS. Ali, KB. Sinha, eds. (Hindustan Book Agency, New Delhi, 2007), pp. 147–160.

\bibitem{GiuggioliJPA2009} L. Giuggioli, F.J. Sevilla, and V.M. Kenkre, 
J. Phys. A: Math. and Theor. {\bf 42}, 434004 (2009).

\bibitem{SevillaPRE2020} F.J. Sevilla, 
Phys. Rev. {\bf E 101}, 022608 (2020).

\bibitem{RomanczukEPJST2012} P. Romanczuk, M. B\"ar, W. Ebeling, B. Lindner, and L. Schimansky-Geier, 
Eur. Phys. J. Special Topics {\bf 202}, 1 (2012). 

\bibitem{ThevesBioPhysJ2013} M. Theves, J. Taktikos, V. Zaburdaev, H. Stark, and C. Beta, Biophys. J. {\bf 105}, 1915 (2013).

\end{thebibliography}
\end{document}